\documentclass[a4paper,12pt,nofootinbib,notitlepage,longbibliography]{article}

\pdfoutput=1
\usepackage{jcappub}
\usepackage{amsmath}
\usepackage{amsfonts}
\usepackage{amssymb}
\usepackage{mathrsfs}
\usepackage{graphicx}
\usepackage{color}
\usepackage[utf8]{inputenc}
\usepackage{amsfonts}
\usepackage{graphicx}
\usepackage{booktabs}
\usepackage{multirow}
\usepackage{makecell}
\usepackage{siunitx}
\usepackage{ragged2e}
\usepackage{rotating}
\usepackage{float}
\usepackage[dvipsnames]{xcolor}
\usepackage{xcolor}
\definecolor{darkred}{rgb}{0.5,0,0}
\definecolor{darkblue}{rgb}{0,0,0.5}
\definecolor{firebrick}{rgb}{0.75,0.125,0.125}
\definecolor{darkgreen}{rgb}{0,0.5,0}
\hypersetup{
        colorlinks,
        urlcolor=darkblue,
	    citecolor=darkgreen,
	    linkcolor=firebrick}

\usepackage{orcidlink}
\usepackage{lipsum}
\usepackage[normalem]{ulem}
\usepackage{enumitem}
\usepackage[capitalise]{cleveref}
\usepackage{pifont}
\usepackage{subfigure}
\usepackage{physics}
\usepackage{comment}

\long\def\exclude#1{}

\title{Illuminating Very Heavy Dark Matter in the Earth with Tau Neutrinos}

\author[a,b]{Tianyi Ding}
\author[a,b]{Ali Kheirandish}
\author[c,d,e]{Qinrui Liu}

\emailAdd{dingt2@unlv.nevada.edu}
\emailAdd{ali.kheirandish@unlv.edu}
\emailAdd{qinrui.liu@queensu.ca}

\affiliation[a]{Dept. of Physics \& Astronomy, University of Nevada Las Vegas, NV, 89154}
\affiliation[b]{Nevada Center for Astrophysics (NCfA), University of Nevada Las Vegas, NV, 89154}
\affiliation[c]{Department of Physics, Engineering Physics and Astronomy, Queen’s University, Kingston ON K7L 3N6, Canada}
\affiliation[d]{Arthur B. McDonald Canadian Astroparticle Physics Research Institute,\\ Kingston ON K7L 3N6, Canada}
\affiliation[e]{Perimeter Institute for Theoretical Physics, Waterloo ON N2L 2Y5, Canada}

\abstract{Dark matter accumulates in the center of the Earth as the planet plows through the dark matter halo in the Milky Way. Possible annihilation of dark matter to Standard Model particles can be probed in indirect dark matter searches. Among possible messengers, neutrinos are uniquely ideal as they can escape dense regions. 
Therefore, neutrino telescopes, with their large volume and broad energy exposures, offer new opportunities to search for dark matter signals from the center of the Earth. However, such studies have been restricted to dark matter masses below $\sim$ PeV as the Earth becomes opaque to very-high-energy neutrinos.
In this study, we demonstrate that neutrino telescopes operating at TeV--PeV energies can probe very heavy dark matter particles if they annihilate to tau neutrinos or tau leptons. Here, we report upper limits on the spin-independent dark matter-nucleon cross section for masses between $10^5$ GeV and $10^{10}$ GeV by using 7.5 years of IceCube high-energy starting event observations. Our results motivate detailed analyses in IceCube and other upcoming neutrino telescopes in the Northern Hemisphere.}
\begin{document}
 \maketitle


\section{Introduction} \label{sec:intro}
Dark matter (DM), an elusive and non-luminous form of matter, is believed to account for approximately 27\% of the Universe's energy density and about 85\% of its matter content~\cite{Planck:2018vyg,Bertone:2004pz}.
A wealth of evidence from cosmological and astrophysical observations—from galaxy rotation curves to gravitational lensing and cosmic microwave background measurements—strongly supports the presence of DM~\cite{Rubin:1980zd, Clowe:2006eq, WMAP:2003elm}.
Yet, its non-gravitational signatures remain undetected despite decades of intensive investigations. Among the many DM candidates, weakly interacting massive particles (WIMPs) have garnered considerable attention because they yield a relic abundance consistent with current cosmological observations~\cite{Roszkowski:2017nbc, Schumann:2019eaa}. 
Their predicted interactions — mediated by the weak nuclear force — provide promising avenues for both direct and indirect detection experiments.
In direct DM detection searches, DM particles are expected to interact with either atomic nuclei or electrons, producing scintillation signals from the recoils~\cite{Goodman:1984dc}. 
To date, the most stringent constraints are provided by experiments such as XENONnT~\cite{XENON:2023cxc}, PandaX~\cite{PANDA-X:2024dlo}, and LUX-ZEPLIN (LZ)~\cite{LZ:2024zvo}. Indirect detection strategies, on the other hand, aim to identify the products of DM annihilation (or decay) to Standard Model particles, which yield observables in the form of stable particles: photons, neutrinos, and cosmic rays~\cite{Silk:1985ax, Gaskins:2016cha, Bergstrom:1998eh, Bergstrom:1988jt}. These indirect searches are primarily focused on regions with high DM densities, such as the Galactic center and nearby dwarf spheroidal galaxies, which are among the most extensively studied targets~\cite{McDaniel:2023bju, McDaniel:2018vam, DiMauro:2019frs}.

Among indirect DM search strategies, neutrino observatories offer a unique window into DM annihilation. Unlike photons or charged particles, 
the weakly interacting nature of neutrinos allows them to escape dense environments and propagate unimpeded through celestial bodies and across cosmological distances.
A trove of data ranging from a few keV to beyond PeV energies is available to the neutrino experiments, enabling indirect search for DM annihilation or decay with solar, diffuse supernova background, atmospheric, and high-energy cosmic neutrinos, see Refs.~\cite{Arguelles:2019ouk, Arguelles:2022nbl} for a recent summary. 
These limits provide competitive constraints, especially at very high energies. In the meantime, there is a distinct possibility that neutrinos are the portal to DM~\cite{Blennow:2019fhy}. As such, DM searches with neutrinos can offer essential clues for ``scotogenic'' models that suggest neutrinos gain mass via interaction with the dark sector~\cite{Boehm:2006mi,Farzan:2012sa,Escudero:2016tzx,Escudero:2016ksa,Hagedorn:2018spx,Alvey:2019jzx,Patel:2019zky, Baumholzer19}.

Celestial bodies such as the Sun and planets capture DM through scattering with ordinary matter~\cite{Gould:1987ir, Nunez-Castineyra:2019odi, Leane:2023woh, Bose:2022ola, Li:2022wix}. Over time, these particles lose sufficient energy to become gravitationally bound, accumulating in high-density regions. If DM particles are Majorana in nature or capable of self-annihilation, their accumulation can result in annihilation processes, producing SM particles~\cite{Gondolo:1990dk}. Compared to other celestial bodies, due to Earth's relatively lower density, neutrinos produced by DM annihilation at its core undergo less attenuation, which makes them detectable by detectors at Earth's surface. 

Neutrino telescopes such as Super-Kamiokande and IceCube are well-suited to search for these annihilation signatures, as they are sensitive to the resulting high-energy neutrino flux at the Earth's surface. Several searches specifically targeting Earth-captured DM have already been carried out by IceCube~\cite{IceCube:2016aga, IceCube:2024yaw}, ANTARES~\cite{ANTARES:2016bxz}, and Super-Kamiokande~\cite{Mijakowski:2018sxg}, each probing potential annihilation signatures in the Earth’s center with no significant excess observed. These studies place constraints on the spin-independent (SI) DM–nucleon cross-section, $\sigma^{\text{SI}}_{\text{N}\chi}$ for various DM annihilation channels, particularly for DM mass less than 10 TeV. Other studies relying on IceCube DeepCore data and assuming an optically thick capturing regime, extend the constraints for DM cross-section to PeV~\cite{Pospelov:2023mlz}.

\begin{figure}[t]
    \centering
    \includegraphics[width=0.85\textwidth]{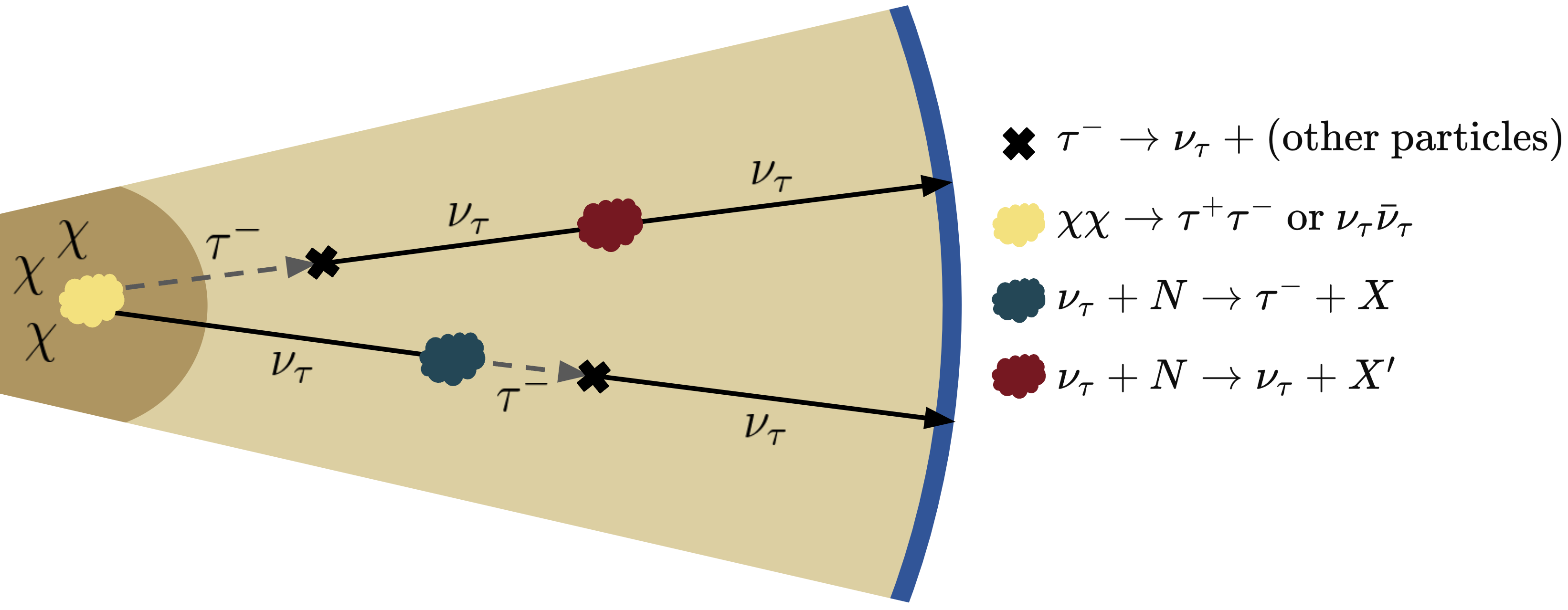}
    \caption{Schematic of \(\nu_\tau\) or \(\tau\) propagation through the Earth. DM annihilation is assumed to occur at the Earth’s center (yellow), producing  \(\nu_\tau\) or \(\tau\) fluxes that propagate toward the Earth’s surface. As shown in the top path, the \(\tau\) lepton can “survive” by decaying into \(\nu_\tau\)(indicated by the black cross) before interacting with Earth. The bottom path illustrates \(\nu_\tau\) continuing to travel outward, undergoing possible charged-current (blue) or neutral-current (red) interactions. Both CC and NC interactions can occur in either the top or bottom path. In the propagation simulation, a 3~km water layer is assumed(blue edge).
    }
    \label{fig: propagation}
\end{figure}

 Indirect searches with neutrino telescopes are mostly limited to $m_\chi$ less than PeV as the Earth becomes opaque to neutrinos at very high energies.
 However, the tau regeneration phenomenon~\cite{Halzen:1998be,Learned:1994wg,Beacom:2001xn,Dutta:2002zc,Bugaev:2003sw} is able to expand our view to larger masses. As shown in Figure~\ref{fig: propagation}, high-energy $\tau$s produced in charged-current $\nu_\tau$ interactions decay into neutrinos, thus the regeneration through $\nu_\tau \leftrightarrow \tau$ conversion leads to an observable flux in a neutrino telescope at lower energies. This cascading process significantly modifies the flux of $\tau$ and $\nu_\tau$ at the Earth's surface, which enhances the potential for detection. In this work, we focus on heavy DM with mass 100~TeV--10~EeV and constrain the spin-independent DM-nucleon scattering cross-section, $\sigma^{\text{SI}}_{\text{N}\chi}$. We consider specific annihilation channels where the tau regeneration effect is particularly crucial -  DM annihilates predominantly into $\tau^+\tau^-$ and $\nu_\tau \bar{\nu}_\tau$. 
 
 This article is organized as follows. In Sec.~\ref{sec:DMann} and Sec.~\ref{sec:propagation}, we discuss the capture and annihilation of DM in Earth and the propagation of the neutrino flux from the Earth's core towards the detector. We present the analysis in Sec.~\ref{sec:analysis} and results in Sec.~\ref{sec:result}. Finally, we conclude in Sec.~\ref{sec:conclusion}. 

\section{Dark matter annihilation}
\label{sec:DMann}
\begin{figure}[t!]
    \centering
    \includegraphics[width=0.8\textwidth]{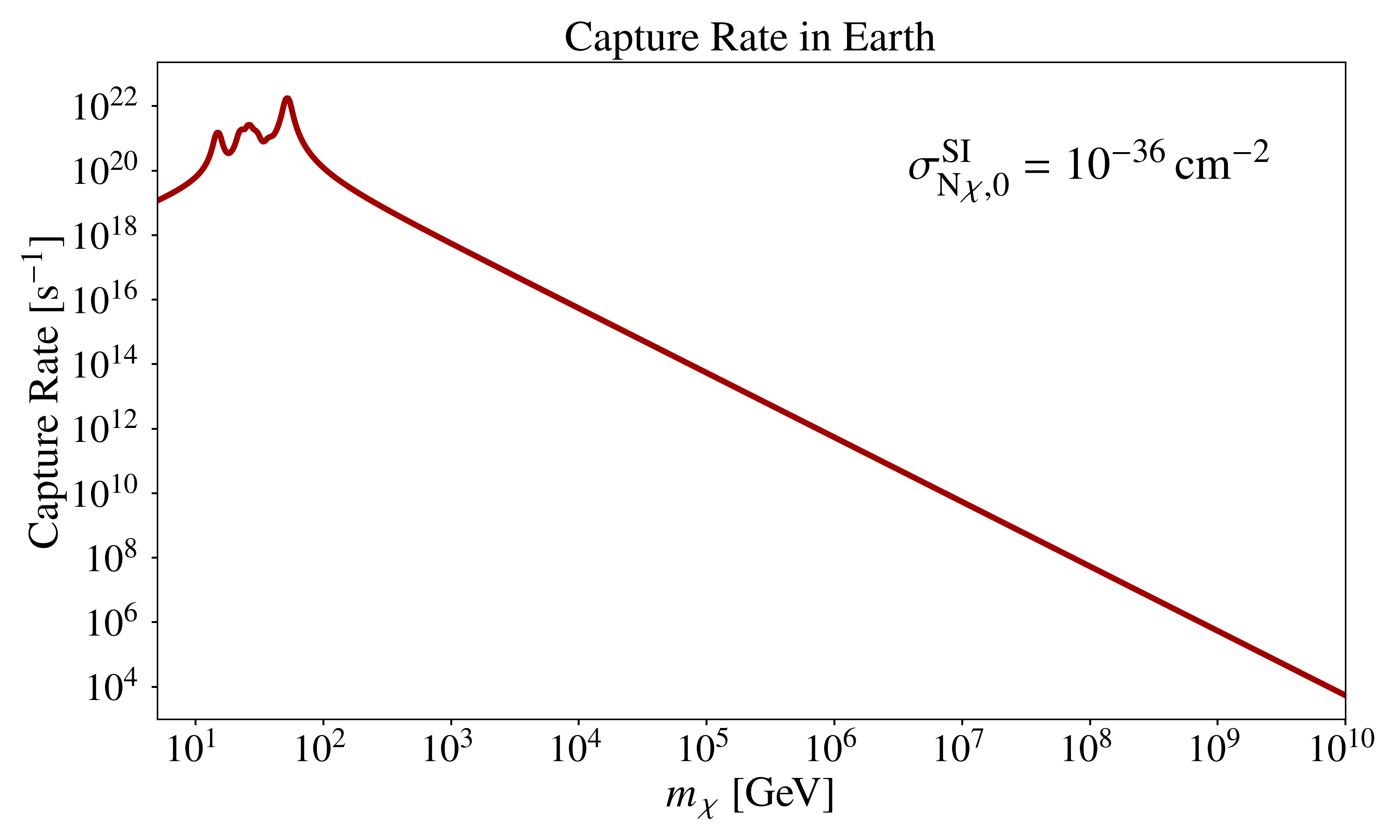}
    \caption{Capture rate as a function of the DM particle mass for the spin-independent DM-nucleon scattering cross section value $\sigma_{\text{N}\chi,0}^{\mathrm{SI}} = 10^{-36} \, \mathrm{cm}^2$. The peaks in the capture rate correspond to resonance capture with the most abundant elements on Earth, namely O, Mg, Si, and Fe.}
    \label{fig: capture_rate}
\end{figure}
\begin{figure}[t]
    \centering
    \includegraphics[width=0.95\textwidth]{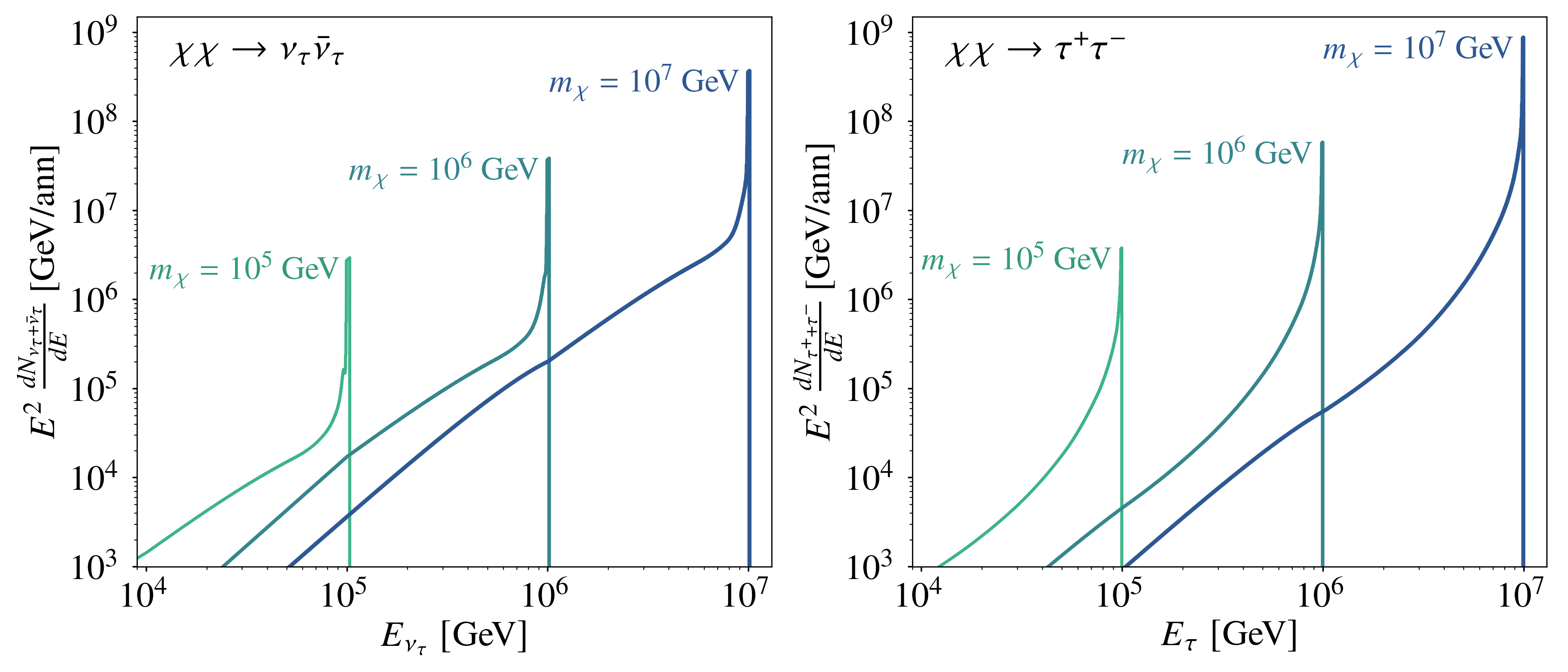}
    \caption{\texttt{\(\mathbf{\chi}\)aro\(\mathbf{\nu}\)}'s predicted ($\nu_\tau + \bar{\nu}_\tau$) flux(left) and ($\tau^+ + \tau^-$) flux(right) per annihilation, assuming two possible DM annihilation channel.  The labels near each peak denote the assumed DM particle mass. This initial flux serves as the input for  near each peak denote the assumed DM particle mass. This initial flux serves as the input for propagation simulations conducted with \texttt{TauRunner}, where it is injected at the Earth's center.}
    \label{fig: initialflux}
\end{figure}

During the capture process, DM particles lose energy until their velocity drops below the gravitational escape threshold. While the DM accumulates in Earth's core, a possible self-annihilation process can reduce the number of DM particles, yielding Standard Model particles. When considering the DM particles heavier than a few GeV inside Earth, the effect of evaporation is negligible~\cite{Gould:1987ju,Garani:2021feo}. Therefore, the temporal evolution of the DM particle population, $N(t)$, can be described by
\begin{equation}
    \frac{dN(t)}{dt} = C - AN(t)^2
\end{equation}
where $C$ denotes the capture rate and $A$ is the annihilation factor. The annihilation rate, $\Gamma_a$, specifying the number of DM annihilation per unit time, is
\begin{equation}
    \Gamma_a = \frac{1}{2}AN(t)^2=\frac{C}{2} \tanh^2(\frac{t}{\sqrt{CA}}).
\end{equation}
Here, $t$ is the time since the Earth's formation. Following Ref.~\cite{Berezinsky:1996ga}, we define $A$ as 
\begin{equation}
A = \frac{\langle \sigma v \rangle}{V_0} \left(\frac{m_\chi}{20 \, \text{GeV}}\right)^{\frac{3}{2}},
\end{equation}
where $m_\chi$ is the DM mass and $\langle \sigma v \rangle$ is the thermally averaged annihilation cross-section rate. The observed DM relic abundance today can be explained if its thermally averaged self-annihilation rate is   
\(\langle \sigma v \rangle \simeq 3 \times 10^{-26} \, \mathrm{cm}^3/\mathrm{s}\) regardless of the annihilation channel~\cite{Jungman:1995df}. Finally, $V_0$ is the effective volume of the Earth, and as derived in~\cite{Berezinsky:1996ga}, that is given by 
\begin{equation}
    V_0 = \left(\frac{3 m_{\text{Pl}}^2 T}{2\rho \times 10 \ \mathrm{GeV}}\right)^{3/2},
\end{equation}
where $T$ and $\rho$ are the central temperature and the central density of the celestial body, and $m_{\text{Pl}}$ is the Planck mass. Thus, for the Earth core, where $T = 6000 \ \mathrm{K}$ and $\rho = 13 \ \mathrm{g} \cdot \mathrm{cm}^{-3}$, the effective volume is $V_0 = 2.3 \times 10^{25} \ \mathrm{cm}^3$~\cite{Dziewonski:1981xy, Balugani:2024ultrafast}.

When DM moves through an object, it scatters with the material and loses energy. DM is captured when the kinetic energy is lower than the gravitational potential and capture can happen via single or multiple scatters~\cite{Press:1985ug,Gould:1987ir,Kouvaris:2010vv,Bramante:2017xlb,Dasgupta:2019juq,Ilie:2020vec,Bramante:2022pmn,Leane:2023woh}. For the DM-nucleon cross section being discussed in this work, the scatterings fall into the single scatter regime as the capture is well below the geometric capture limit corresponding to the case that all DM is captured. The total capture rate has 
\begin{equation}
  C_\oplus=\sum_i C_i,  
\end{equation}
where $C_i$ is the capture rate by element $i$ in Earth which can be written as~\cite{Gould:1987ir,Jungman:1995df}

\begin{align}\label{eq:WIMP_capture}
\begin{split}
    C_i &=  
    \int_0^{R_\oplus} 4\pi r^2 dr 
    \int_0^\infty du \frac{f(u)}{u}\omega(u,r) \Omega^-_i(\omega) \\
    & = \sqrt{\frac{6}{\pi}}\frac{\rho_\chi\sigma_{\chi i}}{{v_\chi} m_\chi}\int_0^R dr\, 4\pi r^2 \frac{\rho_i(r)}{m_i}v^2_\mathrm{esc}(r)\left[ 1-\frac{1-e^{-\mathcal{K}^2_i(r)}}{\mathcal{K}^2_i(r)}\right],
\end{split}
\end{align}
where $R_\oplus$ is Earth radius, $\rho_\chi=0.42~\rm{GeV/cm^3}$ is the local DM density, $u$ is the DM velocity far from the object, $f(u)$ is the velocity distribution and $\omega =\sqrt{u^2+v^2_\mathrm{esc}(r)}$ is the DM speed at $r$ with $v_\mathrm{esc}$ being the escape velocity at the same location. $v_\chi=270~\rm{km/s}$ is the velocity dispersion and the exponential factor $\mathcal{K}^2_i(r) =\frac{3}{2}\left (\frac{v_\mathrm{esc}}{\bar{v}}\right )^2\frac{2m_\chi m_i}{(m_\chi-m_i)^2}$. $\Omega^-_i $ represents the rate of scattering
\begin{equation}
    \Omega_{i}^{-}=\frac{\sigma_{\chi i} \rho_{i}(r)}{m_i} \frac{4 \mu_{+}^{2}}{\mu w} \int_{\frac{w\left|\mu_{-}\right|}{\mu_{+}}}^{v_\mathrm{esc}} dv v\left|F_{i}\right|^{2},
\end{equation}
where $\mu=m_\chi/m_i$, $\mu_\pm=(\mu\pm 1)/2$ and $F_i$ is the nuclear form factor which we calculate following Ref.~\cite{Jungman:1995df}. Since the main elements in Earth predominantly lack intrinsic spin, the most relevant scattering process is spin-independent dark matter-nucleon scattering, $\sigma^{\text{SI}}_{\text{N}\chi}$ and for each element there is $\sigma_{\chi_{i}}=\sigma^{\rm{SI}}_{N\chi}A^2_i\left (\frac{m_i}{m_N} \right)^2 \left (\frac{m_\chi+m_N}{m_\chi + m_i} \right)^2$ where $A_i$ is the atomic mass number. Figure~\ref{fig: capture_rate} shows the Earth’s capture rate as a function of the DM mass. The peaks in the capture rate arise from resonant capture with the most abundant elements on Earth.

Once DM particles are gravitationally captured in the Earth’s core, they accumulate and eventually annihilate, producing high‑energy SM particles. This study concentrates on the annihilation into
$\tau^{+}\tau^{-}$ or $\nu_{\tau}\bar{\nu}_{\tau}$,
because the resulting $\tau$‑leptons and $\nu_{\tau}$ neutrinos propagate
through the Earth with a comparatively higher probability of generating
detectable signals(see Sec.~\ref{sec:propagation}). The initial $\tau$ and $\nu_\tau$ fluxes originating from very heavy DM annihilations in the Earth's center are generated using the \texttt{\(\mathbf{\chi}\)aro\(\mathbf{\nu}\)} software package which is a software organizing calculations of neutrino fluxes from DM annihilation/decay in various environments including Halos and celestial bodies from production to detection~\cite{Liu:2020ckq}. For the $\tau^+\tau^-$ channel, in order to generate the initial $\tau$ flux we need to modify current \texttt{\(\mathbf{\chi}\)aro\(\mathbf{\nu}\)} by turning off the decay of $\tau$. The pre-generated fluxes for DM mass from 500~GeV to 10 TeV incorporate the electroweak corrections calculated for the \texttt{HDMSpectra}~\cite{Bauer:2020jay}. However, the intermediate results of this feature needed here is not employed in \texttt{\(\mathbf{\chi}\)aro\(\mathbf{\nu}\)} itself, therefore, we only generate the fluxes with electroweak showering included in \texttt{pythia8.2}~\cite{Sjostrand:2014zea} for this work.
We consider DM masses from $10^5$~GeV to $10^{10}$~GeV, annihilating through both the $\tau^+\tau^-$ and $\nu_\tau \bar{\nu}_\tau$ channels.
Figure~\ref{fig: initialflux} shows the differential fluxes at production. Then, the flux provided by \texttt{\(\mathbf{\chi}\)aro\(\mathbf{\nu}\)} is converted into a cumulative distribution function (CDF) to serve as the initial flux input for \texttt{TauRunner}~\cite{Safa:2021ghs}, which simulates neutrino propagation through the Earth.
\section{Propagation Through Earth}
\label{sec:propagation}
\begin{figure}[t]
    \centering
    \includegraphics[width=1\textwidth]{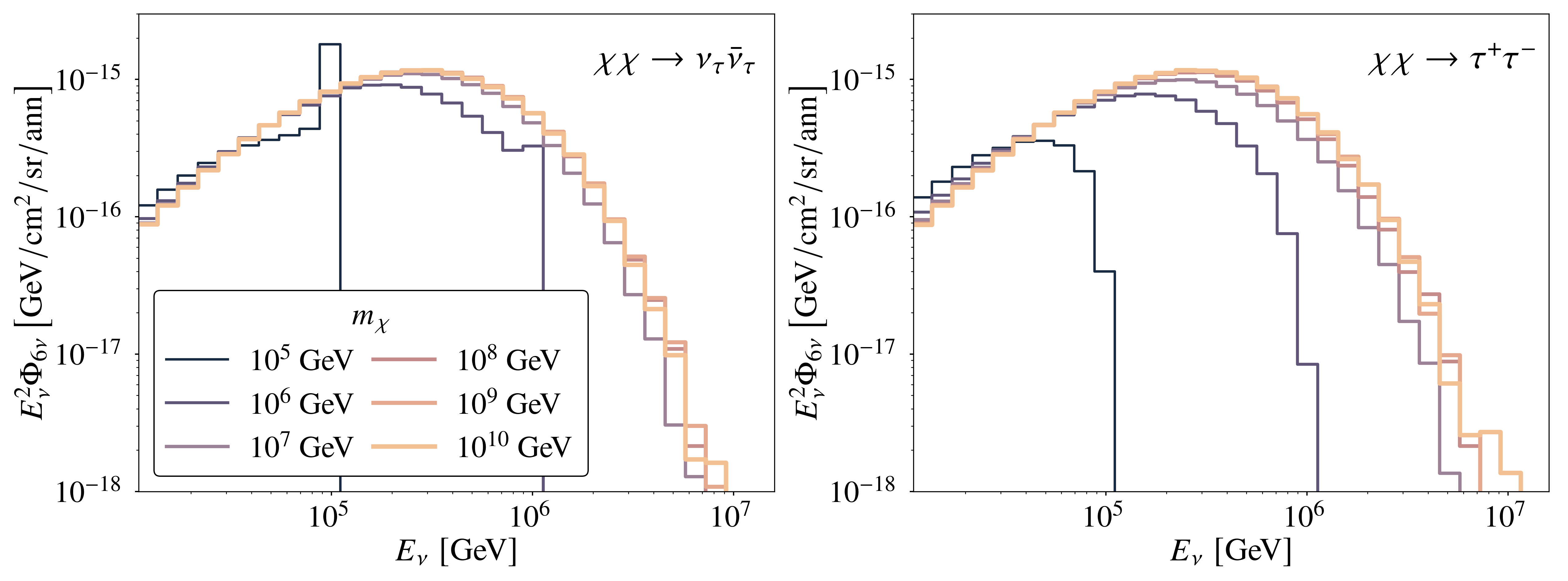}
    \caption{Simulated energy spectra at the Earth surface of all flavor neutrino differential flux per DM annihilation for each DM mass. The left panel is the result assuming $\tau^+\tau^-$ annihilation channel, while the right panel is the result from $\nu_\tau \bar{\nu}_\tau$ annihilation channel.}
    \label{fig: surflux}
\end{figure}
In this study, the initial flux of \(\tau\) and \(\nu_\tau\) from DM annihilation is assumed to originate at the center of the Earth. These high-energy particles propagate through the dense Earth's medium, undergoing a series of interactions and decays that govern the flux observed at the Earth's surface. The physical mechanisms involved in this process are described by the transport equation, which models the attenuation, energy loss, and regeneration of $\nu_\tau$ and $\tau$ during propagation.
The transport equation takes the following form:
\begin{equation}
\frac{d \vec{\phi}(E, x)}{dx} = -\sigma(E) \vec{\phi}(E, x) + \int_E^\infty d\tilde{E} \, f(\tilde{E}, E) \vec{\phi}(\tilde{E}, x),
\end{equation}
where \( E \) is the energy of the particle, \( x \) is the column density, \( \sigma(E) \) is the deep-inelastic scattering cross-section per target nucleon, and \( f(\tilde{E}, E) \) represents the redistribution function from higher energy \(\tilde{E}\) to lower energy \(E\). The spectrum vector \( \vec{\phi}(E, x) \) includes the tau neutrino flux and its antiparticle counterpart. The first term on the right-hand side represents the attenuation of flux at energy \(E\) due to charged-current (CC) and neutral-current (NC) interactions. In contrast, the second term captures the sum of contributions of neutrinos with higher energies cascading down to \(E\) through NC interactions for all-flavor neutrinos and neutrinos from the decay of $\tau$ produced from CC interactions specific to \(\nu_\tau\).

Here, the key feature is the CC interaction of \(\nu_\tau\) propagation, where a \(\nu_\tau\) interacts with a nucleon (\(\nu_\tau + N \to \tau + X\)) to produce a \(\tau\). Charged leptons lose energy in dense media through many processes based on their energy, such as ionization, bremsstrahlung, electron-positron pairs, and photonuclear interaction. However, because \(\tau\) is short-lived with a rest-frame lifetime of approximately \( 10^{-13} \, \text{s} \)~\cite{ALEPH:1997roz}, the produced \(\tau\) has a much higher probability of decaying into high-energy neutrinos before losing energy rapidly. Thus, the primary \(\nu_\tau\) flux is not attenuated but cascades down to lower energies. This process is demonstrated in Figure~\ref{fig: propagation}. By this tau regeneration process, the cascade produced by \(\nu_\tau\) or \(\tau\) flux generated in the center of the Earth has a chance to be detected at the Earth's surface.

To accurately account for these complex processes, we employ the \texttt{TauRunner} package. This effect is also implemented in various neutrino propagation code such as \texttt{nuFATE}~\cite{Vincent:2017svp}, \texttt{NuTauSim}~\cite{Alvarez-Muniz:2018owm}, \texttt{NuPropEarth}~\cite{Garcia:2020jwr}, \texttt{nuPyProp}~\cite{Garg:2022ugd}. The \texttt{TauRunner} software utilizes Monte Carlo (MC) methods to propagate leptons and neutrinos over an energy range from 10\,GeV to \(10^{12}\)\,GeV. Secondary neutrino production arising from charged-current \(\nu_\tau\) interactions is included, and the energy loss of the resulting charged tau is treated stochastically. \texttt{TauRunner} adopts the layered density profile from the Preliminary Reference Earth Model (PREM)~\cite{Dziewonski:1981xy}, and we also include a 3~km water layer representing the ice sheet thickness at the South Pole.
\begin{figure}[t]
    \centering
    \includegraphics[width=1\textwidth]{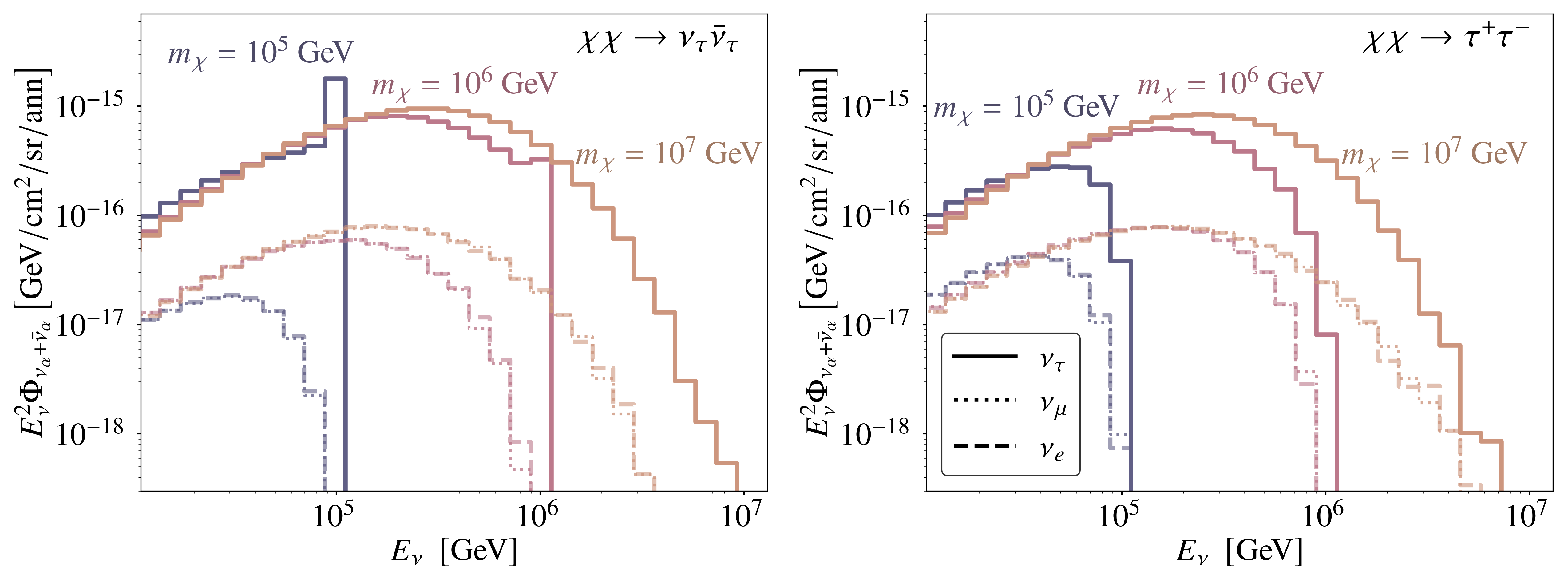}
    \caption{Comparison of the expected surface flux of each neutrino flavor per annihilation. The left panel assumes DM annihilates into $\nu_\tau\bar{\nu}_\tau$, while the right panel is for annihilation into $\tau^+\tau^-$. The line color indicates the DM particle mass, $m_\chi = 10^5, 10^6, 10^{7}$ GeV. Different line styles mark the neutrino flavors. In both cases, $\nu_\tau$ flux dominates over the other two flavors.}
    \label{fig: compare_flavor}
\end{figure}
Figure~\ref{fig: surflux} presents the resulting energy spectra at the Earth's surface for the differential all-flavor neutrino flux per dark matter annihilation. This flux is dominated by $\nu_\tau$, as shown in Figure~\ref{fig: compare_flavor}. 
$\nu_e$ and $\nu_\mu$ can also be produced in $\tau$ decays, however, there is no regeneration effect for them and the flux gets absorbed more rapidly. For $m_{\chi} > 10^7\,$GeV, both the $\tau^+\tau^-$ and $\nu_\tau \bar{\nu}_\tau$ channels exhibit nearly identical fluxes at the surface. However, at lower masses ($m_{\chi} = 10^5, 10^6, 10^7$ GeV), the flux in the $\tau^+\tau^-$ channel undergoes more pronounced attenuation. In this regime, the $\tau$ leptons have a larger chance to experience photonuclear interactions in the Earth's dense core before they decay into neutrinos, redistributing their energy and leading to a broader but more attenuated flux by the time they emerge at the surface.

\section{Analysis}\label{sec:analysis} 

In this work, we adopt a binned-likelihood test to derive constraints on the DM annihilation rate utilizing the IceCube 7.5yr High-Energy Starting Events (HESE)~\cite{IceCube:2020wum}. This dataset focuses on high-energy neutrino events from all sky with vertices within IceCube's fiducial volume, using an outer veto region to suppress backgrounds from atmospheric muons and neutrinos~\cite{Stachurska:2019srh}. Over the 2,635 days of exposure, 102 events were recorded, with 60 events exceeding the 60~TeV threshold, categorized into cascades, tracks, and double cascades based on their morphologies.      

We restrict our analysis to up-going events with zenith angle $>90^\circ$ as DM annihilation happens inside Earth. The overwhelming atmospheric muons are suppressed for up-going events as the Earth acts as a filter to them. 
For deposited energies $>60$ TeV, 21 up-going events are reported in HESE 7.5-year.
Considering our signal expectation, that is dominated by $\nu_\tau$, we exclusively focus on cascade and double-cascade events. This combination is chosen because it provides the most powerful constraint on the dark matter annihilation rate, and further details are provided in Appendix~\ref{morphology_cp_selection}.

Cascades are generated by NC interactions from all neutrino flavors and CC interactions of electron neutrinos, while double-cascade events uniquely signal tau neutrino CC interactions, characterized by two spatially separated energy depositions~\cite{Huang:2018unz}. Notably, when the tau decay length is too short to be resolved, tau neutrino events may also manifest as single cascades. In the 7.5-year HESE dataset, 15 cascade events were observed in the Northern sky sample, but no double-cascade events were detected. Although two double-cascade candidates were identified in the full HESE dataset, both were downgoing events and therefore lie outside the Northern sky selection considered in this analysis~\cite{IceCube:2015rro}.
\begin{figure}[t]
    \centering
    \includegraphics[width=0.8\textwidth]{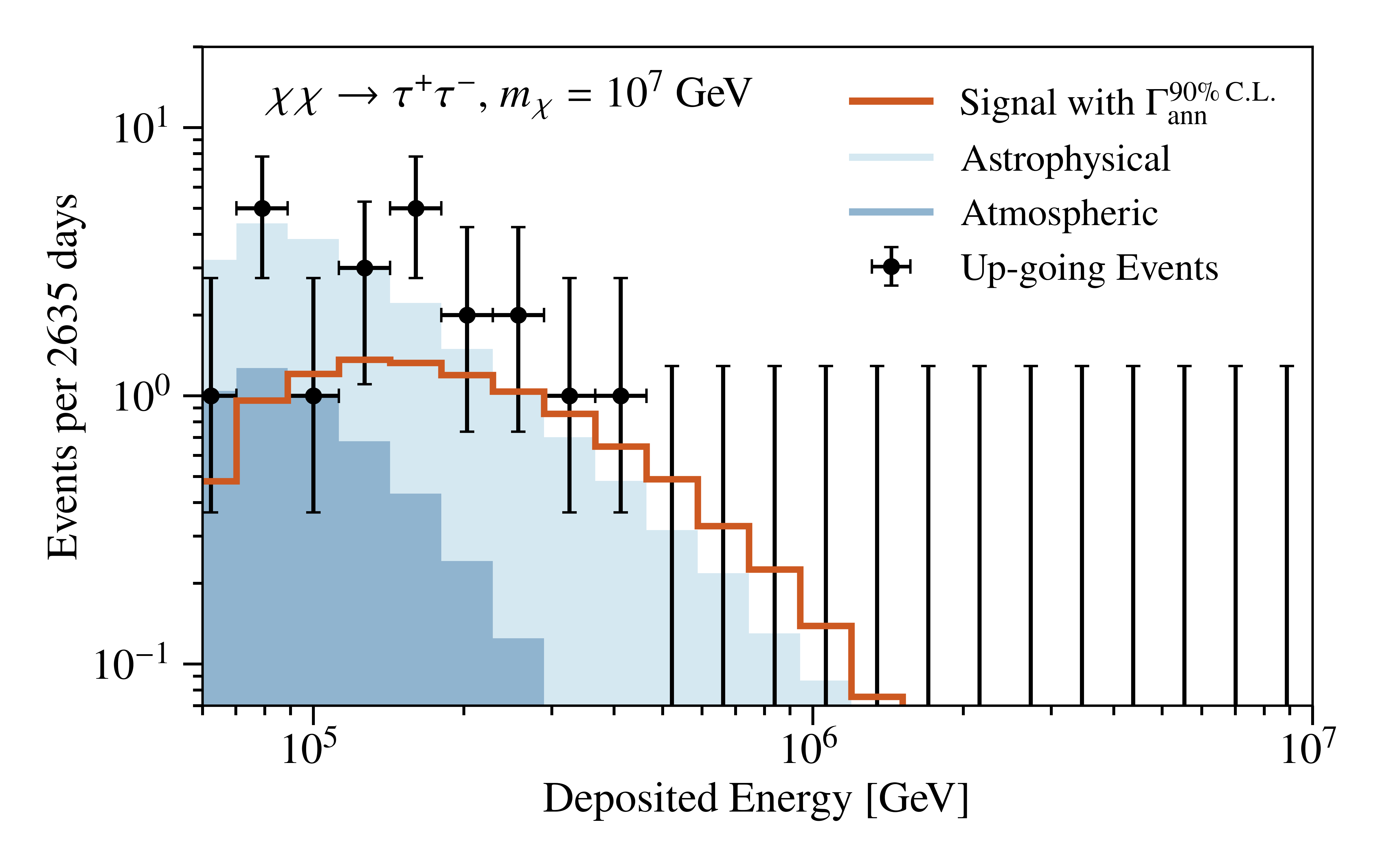}
\caption{The number of expected astrophysical, atmospheric, and DM cascade and double-cascade up-going events compared to observed data in 2635 days of IceCube observation. The light blue histogram represents the expected number of events per flavor based on the astrophysical flux parameterized by a single power-law model with a spectral index \(\gamma_{\text{astro}} = 2.87^{+0.20}_{-0.19}\), as determined from the IceCube 7.5-year HESE analysis. The dark blue histogram indicates the atmospheric background. The black points with error bars show the observed data of up-going events(zenith angle $>90^\circ$), with statistical uncertainties corresponding to the 1$\sigma$ level. The red line is the expected DM signal for $\tau^+\tau^-$ annihilation channel with $m_\chi = 10^7$ GeV, assuming the computed 90\% upper limit of the annihilation rate $\Gamma_{\text{ann}}^{90\% \, \mathrm{C.L.}} = 5.67\times 10^6$ $\rm{s}^{-1}$.}    
    \label{fig: N_envents}
\end{figure}
The background for this analysis includes both astrophysical and atmospheric neutrinos.
For the astrophysical neutrino flux, we assume the best-fit parameters for a single power-law spectrum from IceCube's 7.5-year HESE analysis~\cite{IceCube:2020wum}. As such, we incorporate an isotropic neutrino flux with a flavor composition of \(\nu_e : \nu_\mu : \nu_\tau = 1 : 1 : 1\) at Earth, as expected from neutrino oscillations over cosmic distances\cite{Palladino:2015vna}. The spectrum is parameterized as
\begin{equation}
\frac{dN_{6\nu}^{\text{astro}}(E)}{dE} = 6.37 \left(\frac{E}{100\rm{TeV}}\right)^{-2.87} \cdot 10^{-18} \, \mathrm{GeV}^{-1} \, \mathrm{cm}^{-2} \, \mathrm{s}^{-1} \, \mathrm{sr}^{-1}.
\end{equation}

For the atmospheric component, we utilize the atmospheric model and best-fit parameters reported in the HESE 7.5-year analysis~\cite{IceCube:2020wum}. 

 The expected number of events with specific morphology, \(\lambda^{{k}}_{i,m}\), in each observable bin given the model parameters can be determined by the following equation. 

\begin{equation}
\begin{aligned}
\lambda^{k}_{i,m} &= T_{\rm{live}} \sum_{\alpha} \int_{\Delta E_i} dE_{\mathrm{reco}} \int d\Omega \int dE_\nu\;
A_{\mathrm{eff}}^\alpha(E_\nu, \Omega)\;
\Phi^{k}_{\nu_\alpha + \bar{\nu}_\alpha}(E_\nu, \Omega) \\
&\quad \times R^\alpha_m(E_\nu, \Omega)\;
\mathcal{P}_m(E_{\mathrm{reco}} \mid E_\nu, \Omega, \alpha)
\end{aligned}
\label{eq:event_number_morphology}
\end{equation}

Here, index $k$ denotes the different components in this analysis: astrophysical, atmospheric, and signal from DM, while $\alpha$ indicates the neutrino flavors. $T_{\rm{live}}$ is the operation time of IceCube, and $A^\alpha_{\mathrm{eff}}(E_\nu, \Omega)$ is the effective area for neutrinos of flavor $\alpha$ with true energy $E_\nu$ and direction $\Omega$. The term $R^\alpha_m(E_\nu, \Omega)$ accounts for the probability that a neutrino of flavor $\alpha$ is reconstructed as a specific event morphology $m$, i.e. a track, a cascade, or a double-cascade. The function $\mathcal{P}_m(E_{\mathrm{reco}} \mid E_\nu, \Omega, \alpha)$ describes the detector’s energy response, quantifying the probability density of reconstructing an event with deposited energy $E_{\mathrm{reco}}$ given the true neutrino energy and direction. The total energy range is divided into discrete bins, and in each bin \(i\), we compare the observed number of events \(n_i^\mathrm{obs}\) to the predicted number of events \(\lambda_i\). In this study, we use the public Monte Carlo (MC) simulation for HESE~\cite{IceCube:2020wum}\footnote{\href{https://github.com/icecube/HESE-7-year-data-release}{https://github.com/icecube/HESE-7-year-data-release}} to estimate the number of cascade and double-cascade events of each component. 

We employ a binned-likelihood in this analysis, assuming that the event counts in each energy bin \(i\) follow a Poisson distribution. Therefore, the likelihood function can be written as
\begin{equation}
  \mathcal{L}(\Gamma_\text{ann}) 
  \;=\;
  \prod_{i=1}^{N_\mathrm{bins}}
  \frac{
    \bigl[\lambda_i(\Gamma_\text{ann})\bigr]^{n_i^\mathrm{obs}}
    \,\exp\bigl[-\lambda_i(\Gamma_\text{ann})\bigr]
  }{
    n_i^\mathrm{obs}!
  },
\end{equation}
where \(n_i^\mathrm{obs}\) is the observed event number and \(\lambda_i\) is the expected number of events in bin \(i\). \(i\) runs over all bins $N_\mathrm{bins}$. \(\lambda_i\) is the sum of signal and background, i.e. $\lambda_i(\Gamma_\text{ann})=\lambda^{\rm{DM}}_i(\Gamma_\text{ann})+\lambda^{\rm{bkg}}_i$, where the $\lambda^{\rm{bkg}}$ includes both astrophysical and atmospheric expected number of events.

To calcualte the number of signal events, since we have already taken into account the neutrino propagation effects including tau regeneration when computing the flux at the detector, we employ the down-going selection in HESE MC to compute $\lambda^{\rm{DM}}_i$ to avoid redundant inclusion of the propagation effect as neutrino absorption is negligible for down-going events. For the background estimation, we directly use the effective area for up-going events. 

We follow the standard likelihood-ratio approach in high-energy physics~\cite{Cowan:2010js, ParticleDataGroup:2022pth}, forming a test statistic (TS) by comparing the background-only (null) hypothesis to the signal-plus-background hypothesis as
\begin{equation}
  \mathrm{TS}(\Gamma_\text{ann})
  \;=\;
  -2
  \Bigl[
    \ln \mathcal{L}(\Gamma_\text{ann})
    \;-\;
    \ln \mathcal{L}_0
  \Bigr].
\end{equation}
 
Under the regularity conditions of Wilks’s theorem, and given that our likelihood function is smooth, continuous, and well‑defined across the relevant parameter space, TS asymptotically follows a \(\chi^2\) distribution with one degree of freedom~\cite{Wilks:1938dza}. We therefore obtain the \(\Gamma_{\text{ann}}^{90\% \, \mathrm{C.L.}}\) by solving  \(\mathrm{TS}(\Gamma_\text{ann}) \ge 2.71\) for the 90th percentile of \(\chi^2\).
The expected number of events corresponding to \(\Gamma_{\text{ann}}^{90\% \, \mathrm{C.L.}}=5.69\times10^6\) \(\rm{s}^{-1}\) is reflected by the red line in Figure~\ref{fig: N_envents}.

\section{Result}\label{sec:result}
\begin{figure}[t]
    \centering
    \includegraphics[width=\textwidth]{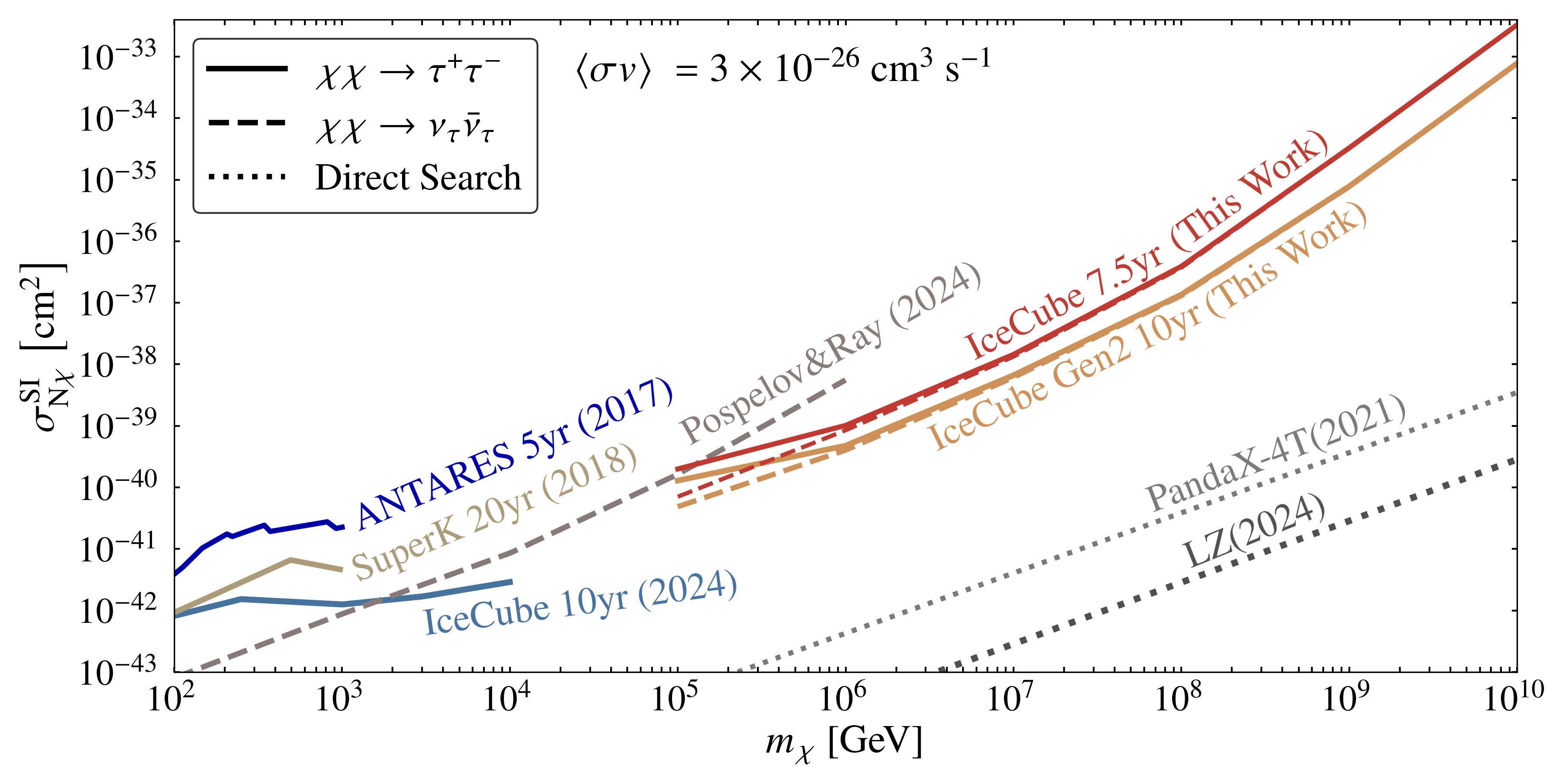}
    \caption{Upper limits on DM-nucleon cross-section $\sigma^{\text{SI}}_{\text{N}\chi}$ and compared to other Earth-bond DM studies, including IceCube~\cite{IceCube:2024yaw}, ANTARES~\cite{ANTARES:2016bxz}, Super-Kaniokande~\cite{Mijakowski:2018sxg}, and Pospelov\&Ray ~\cite{Pospelov:2023mlz}. The IceCube, ANTARES, and Super-Kamiokande results assume DM annihilation into $\tau^+\tau^-$, while the Pospelov\ Ray's limit is derived assuming annihilation into $\nu_\tau\bar{\nu}_\tau$. The solid lines are the upper limits from $\tau^+\tau^-$ channel, while the dashed lines indicate the $\nu_\tau \bar\nu_\tau$ channel result. The red lines denote the result of this study. The expected improvement of upper limits for IceCube Gen2 is shown by the orange line. The gray dotted line is the upper limit by direct DM search, the LUX-Zeplin~\cite{LZ:2024zvo} and PandaX-4T~\cite{PandaX-4T:2021bab}. The thermally averaged self-annihilation rate used in this calculation is \(\langle \sigma v \rangle = 3 \times 10^{-26} \, \mathrm{cm}^3/\mathrm{s}\).}    
    \label{fig: upperlimit}
\end{figure}
\begin{figure}[t]
    \centering
    \includegraphics[width=0.9\textwidth]{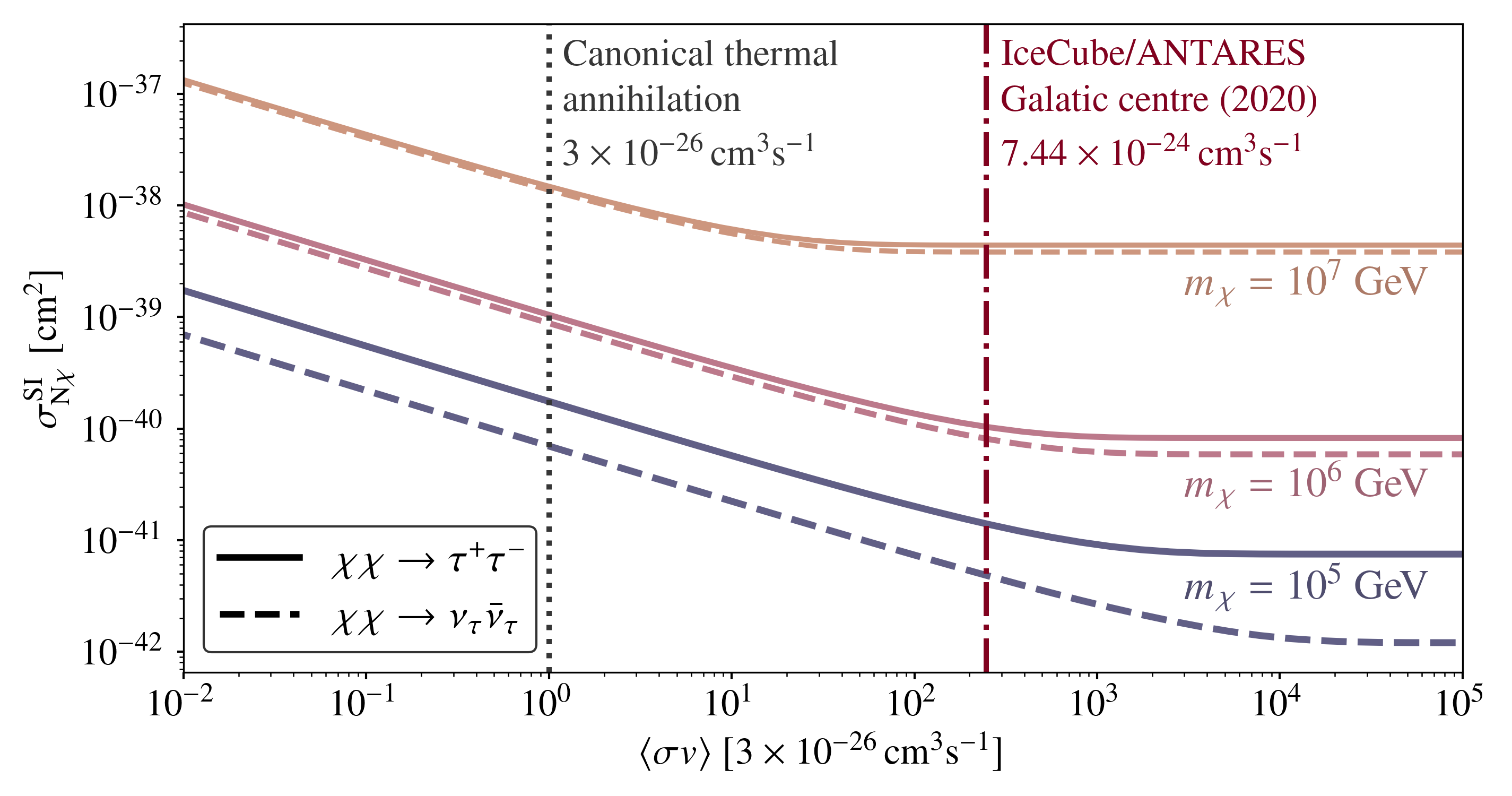}
    \caption{Upper limits on DM-nucleon cross-section $\sigma^{\text{SI}}_{\text{N}\chi}$ as a function of $\langle \sigma v \rangle$.  The solid lines are derived from $\tau^+\tau^-$ channel, while the dashed lines are obtained from $\nu_\tau \bar\nu_\tau$ channel. The corresponding DM masses are denoted at the right end of the lines. The vertical dotted line indicates the canonical thermal annihilation cross-section \(3\times 10^{-26}\,\mathrm{cm^3\,s^{-1}}\),
while the vertical dash-dotted line marks the IceCube/ANTARES limit on the annihilation cross-section from the Galactic Halo \(7.44\times 10^{-24}\,\mathrm{cm^3\,s^{-1}}\)~\cite{ANTARES:2020leh}.}    
    \label{fig: sigma_v}
\end{figure}
For both $\tau^+\tau^-$ and $\nu_\tau \bar{\nu}_\tau$ annihilation channel and DM masses between $10^5$ GeV and $10^{11}$ GeV, we set the 90$\%$ CL upper limits on DM annihilation rate. Subsequently, the limits on the annihilation rate are converted to limits on spin-independent DM-nucleon cross-sections, $\sigma^{\text{SI}}_{\text{N}\chi}$, following equations in Section~\ref{sec:DMann}. 
The red lines in Figure~\ref{fig: upperlimit} show these results for the $\nu_\tau$ and $\tau$ channels. For $m_\chi > 10^7$ GeV, the limits converge. 
On the other hand, for $m_\chi \le 10^7$ GeV, the $\tau^+\tau^-$ channel yields a weaker constraint compared to $\nu_\tau \bar{\nu}_\tau$ channel. This difference arises due to the additional energy losses in the $\tau^+\tau^-$ channel during propagation, as shown in Figure~\ref{fig: surflux}. For comparison, upper limits by other Earth-captured DM  probes are shown in Figure~\ref{fig: upperlimit}. We extrapolate LZ and PandaX-4T's results from $m_\chi$ = $10^{4}$ to $m_\chi$ = $10^{10}$ GeV shown in the gray dotted line.

A broader view of these results can be obtained by mapping the 90\%\,C.L.\ limits onto the 
$\langle \sigma v\rangle - \sigma^{\mathrm{SI}}_{N\chi}$ plane, as shown in Figure~\ref{fig: sigma_v}. For large $\langle \sigma v\rangle$, the capture and annihilation processes within the Earth effectively reach equilibrium, and so the constraint on $\sigma^{\mathrm{SI}}_{N\chi}$ saturates, which results in a horizontal plateau in the plot. As a reference, the dotted vertical line shows the canonical relic cross-section $\langle \sigma v\rangle = 3\times 10^{-26}\,\mathrm{cm^3\,s^{-1}}$, commonly associated with thermal WIMPs. On the other hand, the dash-dotted line marks the combined IceCube and ANTARES limit toward the Galactic center~\cite{ANTARES:2020leh}.

We also examined the prospects for improvement of upper limits with the next generation of IceCube Neutrino Observatory, IceCube-Gen2~\cite{IceCube-Gen2:2020qha}.
The geometric instrumental volume will be increased from 1 $\mathrm{km}^3$ to 7.9 $\mathrm{km}^3$, leading to larger statistics, for which the upper limits are expected to be improved by about 3 times, as shown by the orange line in Figure~\ref{fig: upperlimit}.

We should note that the estimation of IceCube-Gen2 sensitivity is based on a background-only likelihood analysis. Unlike the IceCube limit we present in this work, which uses observed data and compares signal flux against background expectations.
The projected sensitivity for the IceCube-Gen2 relies on simulating the background.
This estimation is tentative, relying on simplified assumptions. It assumes pure volume scaling, where signal and background rates both increase linearly with detector volume. 
In practice, detection efficiency varies with energy, and event selection criteria could influence the actual sensitivity gains. 
Moreover, anticipated enhancements in IceCube-Gen2—such as improved optical sensors, optimized array geometry, and better event reconstruction—are not included here. These advancements could further enhance background rejection and signal detection, potentially yielding even stronger constraints on dark matter annihilation cross-sections beyond those predicted by simple volume scaling.

\section{Discussion \& Summary}\label{sec:conclusion}
In this work, we investigated neutrino signals from DM annihilation within Earth's core and particularly emphasized for the first time the tau regeneration effect, which enables detection of very heavy DM particles (100 TeV--10 EeV). We simulated neutrino propagation inside Earth and compared it with the IceCube 7.5-year High-Energy Starting Events (HESE) dataset to establish the upper limit on the heavy DM annihilation in two major channels. 

Our results show that tau regeneration significantly boosts neutrino signal detectability from DM annihilation, especially at masses above PeV. 
Traditional neutrino searches at these energies typically suffer from severe attenuation due to Earth's opacity. 
By simulating neutrino fluxes for two annihilation channels, $\tau^+ \tau^-$ and $\nu_\tau \bar{\nu}_\tau$, we obtained constraints on the spin-independent DM–nucleon scattering cross-section $\sigma^{\text{SI}}_{\text{N}\chi}$. 
These constraints extend existing limits into DM mass regimes of 10 EeV. For DM masses above $10^7$ GeV, results from the $\tau^+ \tau^-$ and $\nu_\tau \bar{\nu}_\tau$ channels converge. 
At lower masses, differences appear mainly due to stronger attenuation and photonuclear interactions experienced by tau leptons within Earth's core~\cite{Armesto:2007tg}.

This study presents a proof of concept for probing very heavy dark matter annihilation in Earth's core and motivates further analysis by neutrino telescopes. Here, we used the IceCube's HESE sample, which is publicly available. However, this dataset is statistically limited and, due to the large angular uncertainties, does not offer the most sensitive search for an experiment like IceCube. Higher statistics datasets like cascades~\cite{Abbasi:2021ryj, IceCube:2023ame, IceCube:2024csv} or analyses aimed at identifying double cascade~\cite{IceCube:2024nhk} could offer a stronger constraint on DM parameters at these mass regimes. In particular, analyses that can restrict the directionality of the search towards a limited solid angle that embodies the core. Similar improvements can be achieved for the next generation of neutrino telescopes. Here, we presented the projected sensitivity of IceCube-Gen2. Additional telescopes, under development in the Northern Sky, would enhance the total volume exposure for DM searches from the Earth. A similar boost in constraint can be envisioned when a combined analysis is performed on the data from all of these detectors. Based on existing proposals, such as Baikal-GVD~\cite{Aynutdinov:2023ifk}, KM3NeT~\cite{Gozzini:2024uat}, and TRIDENT~\cite{TRIDENT:2022hql}, the total effective volume of neutrino observatories is projected to reach approximately 20~$\rm{km}^3$~\cite{Song:2020nfh}. When combined with the longer operation times of current detectors, such as IceCube, the resulting increase in statistics will significantly enhance the sensitivity to rare signals. This will further advance the capability of neutrino telescopes to look for heavy DM annihilation signatures from the Earth.

\section*{Acknowledgments}

The authors would like to thank Joseph Bramante, Ke Fang, Ningqiang Song, Aaron Vincent, and Tianlu Yuan for their helpful discussion. AK is supported by NASA award 80NSSC23M0104. QL is supported by the Arthur B. McDonald Canadian Astroparticle Physics Research Institute, with equipment funded by the Canada Foundation for Innovation and the Province of Ontario, and housed at the Queen’s Centre for Advanced Computing. Research at Perimeter Institute is supported by the Government of Canada through the Department of Innovation, Science, and Economic Development, and by the Province of Ontario. 

\bibliographystyle{JHEP-2}

\providecommand{\href}[2]{#2}\begingroup\raggedright\endgroup

\newpage
\appendix
\section{Event Morphology Selection}\label{morphology_cp_selection}
\begin{figure}[t]
    \centering
    \includegraphics[width=1\textwidth]{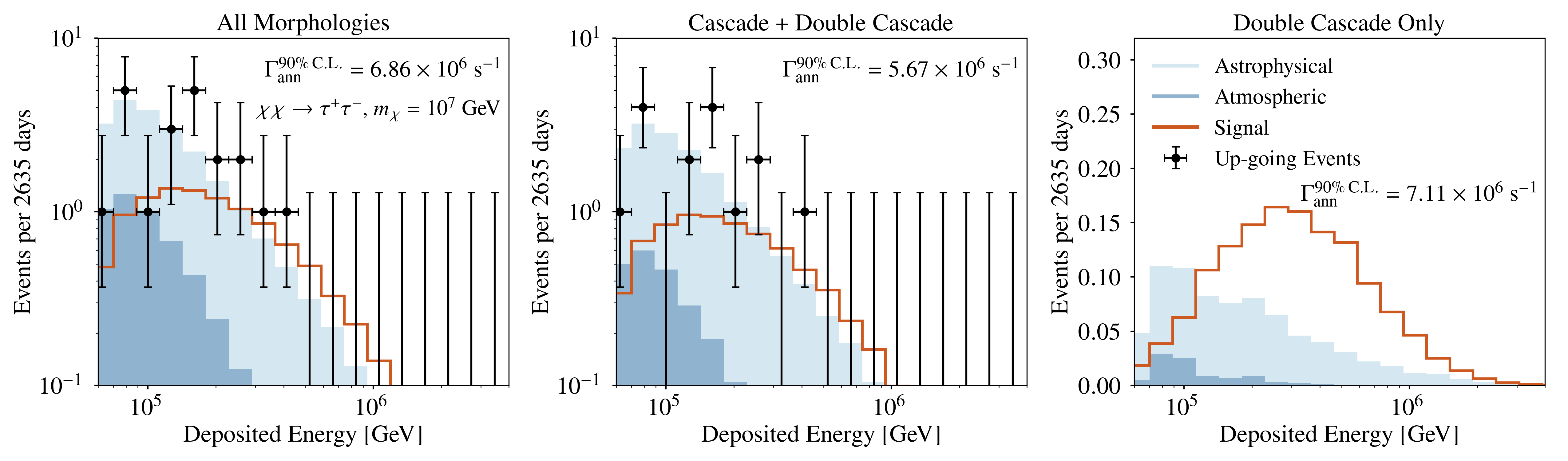}
        \caption{Comparison of the expected dark matter signal and observed IceCube events for different event morphologies. The plots show event distributions for an assumed dark matter annihilation into $\tau^+\tau^-$ at a particle mass of $m_\chi = 10^7\,\text{GeV}$. Three morphology selections are considered: all morphologies——cascade, track, double cascade (left), cascade and double cascade combined (center) , and double cascade only (right). The astrophysical neutrino background (light blue), atmospheric neutrino background (dark blue), observed IceCube data (black points with error bars), and expected dark matter signal at the 90\% confidence-level upper limit (orange line) are shown. With the current reconstruction capability, the Cascade + Double Cascade selection yields the strongest upper limit.}
    \label{morphology_cp}
\end{figure}
In this analysis, we considered three event morphologies reported by the IceCube HESE 7.5yr dataset: cascade, track, and double cascade. Cascade events primarily arise from NC interactions of \(\nu_\tau\), \(\nu_\mu\), and \(\nu_e\), as well as CC interactions of \( \nu_e \), whereas double-cascade events uniquely originate from \(\nu_\tau\) CC interactions. Track events, primarily produced by \(\nu_\mu\) CC interactions, exhibit elongated, track-like signatures that are less distinct for identifying \(\nu_\tau\) signals. In IceCube HESE 7.5-year, there are 21 up-going events with energy \(>60\) TeV. Among them, 15 are cascades, and 6 are tracks, while no double-cascade events were reported from the northern sky.

In our examination, we found that the Cascade plus Double Cascade selection provides the strongest upper limit. As an example, Figure~\ref{morphology_cp} shows the result of \(\Gamma_{\text{ann}}^{90\% \, \mathrm{C.L.}}\) for each case with the event distribution of the observed and background expectation. The upper limit on annihilation rate, \(\Gamma_{\text{ann}}^{90\% \, \mathrm{C.L.}}\), is improved by about 17\% and 20\% compared with using all events and only double-cascade events. This improvement will be slightly less obvious for lower DM masses but will be marginally stronger for heavier DM particles.  

The improvement from the ``All Morphologies" selection to the Cascade plus Double Cascade selection primarily comes from the removal of track events, which are predominantly background, improving the signal-to-background ratio and leading to a stronger upper limit on \(\Gamma_{\text{ann}}^{90\% \, \mathrm{C.L.}}\). Despite double cascades being a unique signature of \(\nu_\tau\) interactions, the lack of observed events in this channel and the small expected background rate result in a weaker statistical constraint. With zero observed double cascades, any predicted DM signal must be compared against a near-zero background, leading to a likelihood test dominated by small numbers of statistics. This limits the statistical power of the constraint. On the other hand, cascade events provide an additional sample where \(\nu_\tau\) interactions can contribute, either through neutral-current interactions or charged-current interactions followed by hadronic or electronic decays. The inclusion of cascades increases the total number of events available for analysis, reducing statistical uncertainties and ultimately yielding the strongest upper limit.

\begin{figure}[t]
    \centering
    \includegraphics[width=0.9\textwidth]{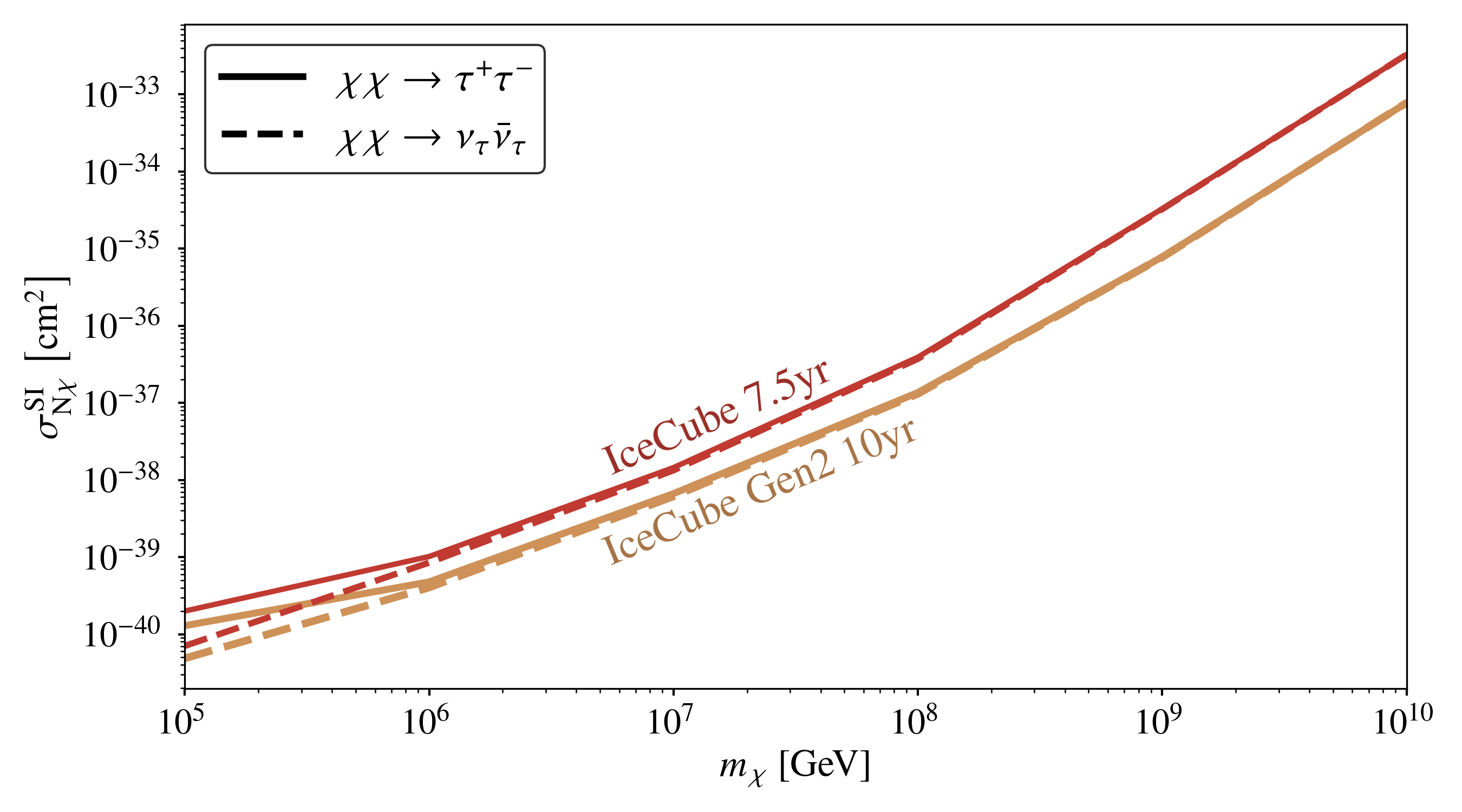}
        \caption{A closer look at the upper limits on the spin-independent dark matter-nucleon cross-section, $\sigma^{\text{SI}}_{\text{N}\chi}$, derived in this work. The solid lines represent the upper limits obtained from the $\tau^+\tau^-$ annihilation channel, while the dashed lines indicate those from the $\nu_\tau \bar\nu_\tau$ annihilation channel. The red lines show the results derived from IceCube 7.5-year data in this study, and the orange lines illustrate the anticipated sensitivity improvements expected from IceCube Gen2 after 10 years of operation.}
    \label{this_work_only}
\end{figure}

For completeness, Figure~\ref{this_work_only} provides a focused view of the limits obtained in this work, displaying only the results based on the IceCube 7.5-year data and the projected IceCube-Gen2 sensitivity.

\end{document}